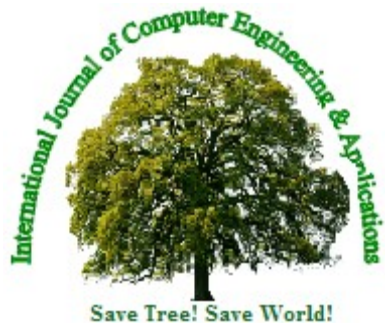

# Towards Securing APIs in Cloud Computing

**Kumar Gunjan [#1], R. K. Tiwari [*2], G. Sahoo [#3]**

[#] *Department of Information Technology, Birla Institute of Technology, Mesra*
*Ranchi, India*
[*] *RVS College of Engineering& Technology*
*Jamshedpur, India*

## ABSTRACT:

Every organisation today wants to adopt cloud computing paradigm and leverage its various advantages. Today everyone is aware of its characteristics which have made it so popular and how it can help the organisations focus on their core activities leaving all IT services development and maintenance to the cloud service providers. But it's the security concerns that customers have which are preventing them to adapt this cloud paradigm whole heartedly, concerns about security of their sensitive data which will lie in the hand of the Cloud Service Providers (CSPs). Application Programming Interfaces (APIs) act as the interface between the CSPs and the consumers. This paper proposes an improved access control mechanism for securing the Cloud APIs.

## [I] INTRODUCTION

The features of Cloud Computing overlap with many other existing technologies such as Grid Computing, Utility Computing, Cluster Computing and Distributed Computing in general. It actually evolved out of Grid Computing and relies on Grid Computing as its backbone and infrastructure support. The evolution has been a result of a shift in focus from an infrastructure that delivers storage and compute resources (such is the case in Grids) to one that is economy based aiming to deliver more abstract resources and services (such is the case in Clouds). As for Utility Computing, it is not a new paradigm of computing infrastructure; rather, it is a business model in which computing resources, such as computation and storage, are packaged as metered services similar to a physical public utility, such as electricity and public switched telephone network. Utility computing is typically implemented using other computing infrastructure (e.g. Grids) with additional accounting and monitoring services. A Cloud infrastructure can be utilized internally by a company or exposed to the public as utility computing [7].

Fig 1: Grids and Clouds Overview

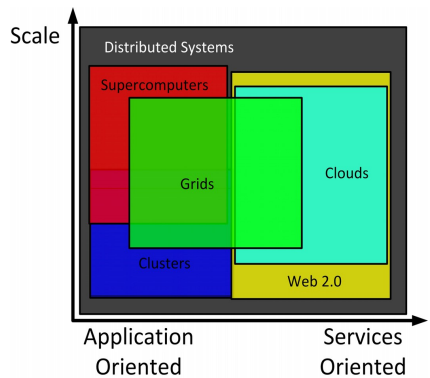

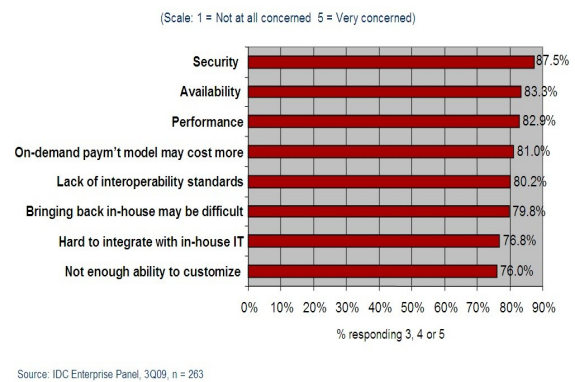

Fig 2: Rating of challenges/issues ascribed to cloud on-demand model

Figure 1, gives an overview of the relationship between Clouds and other domains that it overlaps with. Web 2.0 covers almost the whole spectrum of service-oriented applications, where Cloud Computing lies at the large-scale side. Supercomputing and Cluster Computing have been more focused on traditional non-service applications. Grid Computing overlaps with all these fields where it is generally considered of lesser scale than supercomputers and Clouds.

In contrast to delivering software, platform or infrastructure in the traditional way, where everything stays in- house or under proper physical and logical control of the owner. In Cloud Computing everything is placed in remote data centres which are large storage and server farms and may be located at different geographic locations. Since all of customer's data is kept on the CSP's side, there arises the need for proper security measure and frameworks.Figure1 shows security aspect as the major challenge in adoption of cloud computing as per the review report by IDC Enterprise panel, September, 2009 [1].

In cloud computing the whole gamut of services which are provided, is done via APIs which acts as the interface between the CSP and the consumer. The provisioning of any of the services, their management, orchestration, and monitoring, all are done using these interfaces [15]. So these interfaces must be designed to protect against the various threats to data residing in the cloud environment [2]. It's a person identity which authorizes him and gives rights to access some data or application in the cloud [3].So all measures must be taken through proper access control mechanisms to prevent abuse of identity. Cloud Security alliance mentions Insecure APIs as one of the Top Threats to Cloud Computing [15]. Hence this paper focuses on the authorization and access control aspects of the Cloud API thereby making the residing cloud data more secure subsequently.

Access control of data should be flexible and fine grained depending on the dynamic nature of the cloud. Access control's role is to control and limit the actions or operations in the Cloud systems that are performed by a user on a set of resources. In brief, it enforces the access control policy of the system, and at the same time it prevents the access policy from subversion. Access control in the literature is also referred to as access authorization or simply authorization [5].

A Cloud access control policy can be defined as a Cloud security requirement that specifies how a user may access a specific resource and when. Such a policy can be enforced in a Cloud system through an access control mechanism. The latter is responsible for granting or denying a user access upon a resource. Finally, an access control model can be defined as an abstract container of a collection of access control mechanism implementations, which is capable of preserving support for the reasoning of the system policies through a conceptual framework. The access control model bridges the existing abstraction gap between the mechanism and the policy in a system [4] - [7].

The rest of the paper is organized as follows: We first give an overview about the various access control mechanisms and other related work in section 2, and then we discuss the requirements for access control in the cloud environment in section 3. In section 4, we present the details of the proposed access control model and provide a series of steps simulating the behaviour of the Cloud API while interacting with any user. Finally, section 5 concludes this paper and gives an insight into future scope of work.

## [II] RELATED WORK

Researchers have developed various access control methods to access a resource in computing systems [8].

Access Control Matrix (ACM) characterizes the rights of each subject to every object in a table where each row is a subject, each column is an object, and each entry is the set of access rights [9]. When an organization has a large number of users, storing and managing their related information becomes difficult. Therefore, ACM lacks scalability and also brings performance issues [10].

In Access Control Lists (ACL), list of permissions are attached to each object and specify all the subjects that have access rights for the object. ACL is implemented in modern operating systems and grants or denies access for the specified user or groups of users. In ACL, it is easy to find who has access to an object, but it is difficult to find all access rights of a user. If the organization is large, ACL lacks flexibility where the access control policy changes constantly [9].

Discretionary Access Control (DAC) depends on the discretion of an object's owner who is authorized to control the information resource access. DAC is ownership based and doesn't provide a high degree of security in distributed systems. If the owner provides read access for a file to another user, the user can still copy the contents of the file. Therefore, there is no assurance in the security and the access rights are decided by the owner and do not follow the organization's security standards [11].

In Mandatory Access Control (MAC) a central authority makes decisions about who has access to what information and MAC is widely used in military security. The owner doesn't have permission to change his or her access rights. A labelling mechanism is used to determine the MAC policy. For instance a user with the 'secret' label is not allowed to read a file with the 'top secret' label. Security labelling in MAC is not flexible and is not convenient for task execution [12].

In Role-Based Access Control (RBAC), users are assigned to roles and roles are assigned to access rights or permissions [13]. Figure 2 shows the basic components of RBAC, i.e., user, role, session and permission. Role Hierarchy allows senior roles to inherit roles from junior roles. Being a passive access control model, RBAC fails in capturing dynamic responsibilities of users to support workflows, which need dynamic activation of access rights for certain tasks.

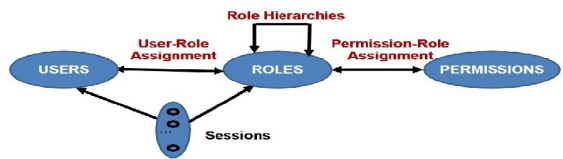

Fig 3: Role based access control model

In Task-Based Authorization Control (TBAC), permissions are activated or deactivated according to the current task or process state [14]. The authorization step is the protection step and with each authorization step the usage count is incremented by one. When the usage count reaches the limit, the corresponding permissions are deactivated. TBAC is an active access control model based on tasks but there is no separation between roles and tasks.

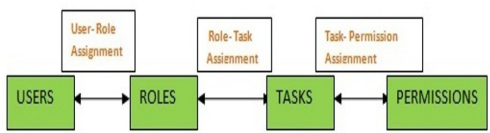

Fig 4: Task role based access control model

In Task-Role-Based Access Control (T-RBAC), users have relationships with permissions through roles and tasks. T-RBAC is an active access control model but delegation of tasks when one user is not available is missing.

## [III] REQUIREMENTS FOR ACCESS CONTROL IN THE CLOUD ENVIRONMENT

The Cloud environment is multi-tenant and heterogeneous environment. To exploit this paradigm and leverage its advantages, CSPs will be providing various services and the same set of services to many customers or tenants as well. Hence the requirements for access control will be different from what we generally do in case of general organisation intranets.

The factors which must be considered in cloud environment are [16]:

A. *Tenant:* For example, the various customers to the CRM application offered by Salesforce.com are the various tenants. For ex MRF Tyres, BIT Mesra, Apollo Hospitals, these organisations can be the customers and each in turn will have a number of users.

B. *User:* The employees of each of the tenants will be considered as users for the various cloud applications.

C. *Task:* The simplest or most basic unit of a business process may be referred to as task.

D. *Objects:* Objects are the various resources which users will like to access*Role:* A role is assigned to each user of any organisation on the basis of various activities which he is allowed to perform.

E. *Permission:* Permission is an authorisation to perform an operation on an object.

F. *Session:* A session maps a user to different roles.

G. *Location:* To incorporate location aware accesses permission with the location. Allowed role locations will also be checked in addition to the normal mapping permissions from role to objects [17].

H. *Business Rules:* Business rules are standard practices of an organization, which its users follow, and may differ from one organization to another. Business rules include:

1) *Least Privilege:* Least Privilege defines that the users are given permissions selectively such that they are not given more permission

than it is necessary to perform their duties [12]. When the users are provided with the ability to perform desired functions, the least privilege policy prevents the issue of a user performing unnecessary and potentially harmful actions.

2) *Least Separation of Duty:* Least Separation of Duty reduces the chance of collusion by distributing the responsibilities for tasks in a business process between multiple users and protects against fraudulent activities of users [18]. Distribution of responsibilities or defining the tasks in workflow or non-workflow could be static, which govern the administration or design time associations between users and permissions, of dynamic, which govern the way in which permissions or task instances are granted at run-time.

3) *Delegation of Tasks:* Delegation of Tasks is also part of business rules and is necessary for access control. It allows performing a task when the initially assigned user is not available to complete the task. The superior role has the rights to reassign the task from one junior role to another and the access rights should be revoked once the task is completed.

### [IV]  PROPOSED WORK

To secure an API the most common trend is to have various authentication techniques like passwords etc to validate the identity of a user and to ensure that the right user is going to access the cloud application or service.

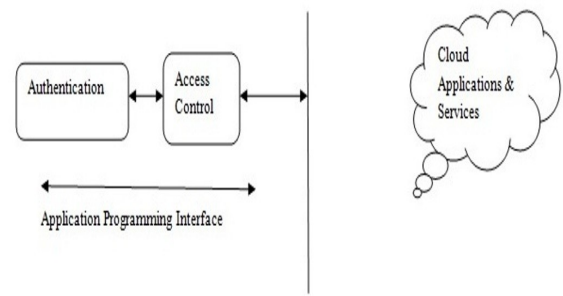

Fig 5: Abstract model of User and Services in the cloud through API

As we know that in case of cloud computing, the provisioning of any of the services, their management, orchestration, and monitoring, all are done using the cloud API [15]. So just validating the fact that the user indeed belongs to the correct tenant or organisation for which this particular service is intended, will not be enough. What if a malicious employee of the right firm, tries to misuse its rights to modify the configuration setting or others features through the cloud API?

So to address this issue we propose using access control policies in addition to the authentication mechanism towards having a secure cloud API. The same has been depicted in Figure 4. Until recently Role based access control models have been very popular since its users-roles-permissions could be mapped with the way our business processes are assigned and work. But in cloud scenario we will need to incorporate few changes in it. It's a passive model and fails to capture the dynamic changes in the responsibilities of the users. We propose to use Task-Role-Based Access Control model in which users have relationships with permissions through roles and tasks. It is inspired from the works suggested in [16] and [17].  This makes the nature of access control mechanism dynamic in nature. Once a user completes his task, his access rights are revoked and is not allowed to access, until he has been granted permission again.

Steps involved when a user requests to access the cloud resources via its API:

a) If the user is new he registers himself at the API by filling its various attributes like name designation, and tenant name. Else go to step (d).

b) The API first checks if the tenant name is correct and also validates the employee name by matching it with say the employee id which will be pre stored in one of the servers at the CSPs side. If the true, go to step (e), else user's request is rejected and the tenant is sent an alert about one unauthorized attempt.

e) If it's a legitimate user then, the API asks the user to enter its request as to what he is interested in performing further. Once the user puts his request, the API reads the user's role and maps it with the tasks he has been granted permission to operate on.

f) If a mapping exists between that role and the task, the user is approved and allowed by the API to perform the task, else an alert message is returned to the user and a similar flagged alert message is mailed to the tenant, giving information about the malicious insider (or employee).

## [V] CONCLUSION

This paper attempts to make Cloud APIs secure which has been considered one of the top threats for cloud computing. The proposed model in addition to the id-validation is done and output is true go to next step. If the output is false user's request is rejected and the tenant is sent an alert about one unauthorized attempt.

c) The user is asked to create and save a password for further validation of his identity in future. Go to step (e).

d) If it's not a new user, the user is asked to enter the password which is validated by the API after computing its hash and matching it with the corresponding value pre-computed and saved in a separate table. If output of authentication is password way of authenticating legitimate users, also exploits the access control policies to prevent unauthorized access to the functionalities provided by the cloud APIs. In future we intend to work on the same model but will try delving further into better authentication and access control models which may be used to provide a better and secured experience to the cloud users.

### ACKNOWLEDGMENT

The authors would like to thank the anonymous reviewers for their valuable suggestions.